\documentclass[twocolumn,twoside,conference,letter]{IEEEtran} 
\usepackage{amsfonts,amsmath,amstext,verbatim}
\usepackage{graphicx}

\def\BibTeX{{\rmfamily B\kern-.05em{\scshape i\kern-.025em b}\kern-.08em \TeX}}

\usepackage{epstopdf}

\graphicspath{{./}%
              {./figures/}}%

\def\EE{\mathbb{E}}

\newcommand\bsection{\section}
\newcommand\bsubsection {\subsection}
\newcommand\bsubsubsection {\subsubsection}

\newcommand{\cref}[1]{Chapter~\ref{#1}}

\newcommand\lame{{\lambda_e}}

\newcommand{\maxarg}{\mathop{\text{\rm arg~max}}}
\newcommand{\dmaxmin}{{d_{maxm}}}

\newcommand{\calL}{\mathcal{L}}

\newcommand{\calS}{\mathcal{S}}

\renewcommand{\Pr}{\mathbf{P}}
\newcommand{\ind}{\mathbf{1}}

\newcommand{\mcd}{\text{\,\rm d}}
\newcommand{\md}{\text{\rm d}}

\newcommand{\ir}{{\mathbb{R}}}
\newcommand{\bbD}{{\mathbb{D}}}
\renewcommand{\EE}{{\mathbf E}}

\newcommand{\othetas}{\overline\theta^*}
\newcommand{\uthetas}{\underline\theta^*}
\newcommand{\obbD}{\overline\bbD}
\newcommand{\ubbD}{\underline\bbD}
\newcommand{\oU}{\overline U}
\newcommand{\uU}{\underline U}
\newcommand{\oprec}{\overline p_{rec}}
\newcommand{\uprec}{\underline p_{rec}}
\newcommand{\odmaxmin}{\overline d_{maxm}}
\newcommand{\udmaxmin}{\underline d_{maxm}}
\newcommand{\oM}{\overline M}
\newcommand{\uM}{\underline M}
\newcommand{\op}{\overline p}
\newcommand{\up}{\underline p}
\newcommand{\orho}{\overline\rho}
\newcommand{\urho}{\underline\rho}
\newcommand{\orhomaxmin}{\overline\rho_{maxm}}
\newcommand{\urhomaxmin}{\underline\rho_{maxm}}
\newcommand{\oRmaxmin}{\overline R_{maxm}}
\newcommand{\uRmaxmin}{\underline R_{maxm}}
\newcommand{\oI}{\overline I}
\newcommand{\uI}{\underline I}
\newcommand{\od}{\overline d}
\newcommand{\ud}{\underline d}
\newcommand{\ods}{\overline d^*}
\newcommand{\uds}{\underline d^*}

\newcommand{\uRs}{\underline R^*}

\newcommand{\resetcounters}{%
        \setcounter{equation}{0}
        \setcounter{Th}{0}}

\renewcommand{\theequation}{\arabic{section}.\arabic{equation}}

\renewcommand\bsection[1]{\section{#1} \resetcounters}

\renewcommand\bsubsection {\subsection }
\renewcommand\bsubsubsection {\subsubsection }

\newtheorem{Th}{Theorem}[section]
\newtheorem{Prop}[Th]{Proposition}

\newtheorem{Cor}[Th]{Corollary}

\newtheorem{ex}[Th]{Example}

\newcommand{\rem}{{\medskip\noindent \em Remark:\ }}

\title{\LARGE On Performance of Event-to-Sink Transport
in Transmit-Only Sensor Networks
\vspace{-2ex}}

\author{
Bart{\l }omiej B{\l }aszczyszyn${}^{1}$ and  
Bo\v zidar Radunovi\'c ${}^{2}$
}

\date{\today}

\begin{document}

\maketitle


\thispagestyle{empty}
\pagestyle{plain}

\renewcommand\thefootnote{}
\footnote{
${}^{1}$~INRIA~\&~ENS
and Mathematical Institute University of Wroc{\l}aw,
45 rue d'Ulm, 75005 Paris FRANCE,
Bartek.Blaszczyszyn@ens.fr;\\
\indent
${}^{2}$~INRIA~\&~ENS
45 rue d'Ulm, 75005 Paris FRANCE,
Bozidar.Radunovic@ens.fr}
\renewcommand\thefootnote{\arabic{footnote}}
\refstepcounter{footnote}

\begin{abstract}
We consider a hybrid wireless sensor network with regular and transmit-only 
sensors. The transmit-only sensors do not have receiver circuit, hence 
are cheaper and less energy consuming, but their transmissions cannot 
be coordinated. Regular sensors, also called cluster-heads, are 
responsible for receiving information from transmit-only sensors and 
forwarding it to sinks. The main goal of such a hybrid network is to 
reduce the cost of deployment while achieving some performance 
constraints (minimum coverage, sensing rate, etc).

In this paper we are interested in the communication between transmit-only 
sensors and cluster-heads. We develop a detailed analytical model of 
the physical and MAC layer using tools from queuing theory and stochastic 
geometry. (The MAC model, that we call Erlang's loss model with
interference, might be of  independent interest as  adequate 
for any non-slotted; i.e.,  unsynchronized, wireless communication channel.)
We give an explicit formula for the frequency of successful 
packet reception by a cluster-head, given sensors' locations. We 
further define packet admission policies at a cluster-head, and we 
calculate the optimal policies for different performance criteria. 
Finally we show that the proposed hybrid network, using the optimal 
policies, can achieve substantial cost savings as compared to 
conventional architectures.
\end{abstract}


\section{Introduction}
\resetcounters


In this paper we analyze performance of a {\em hybrid sensor network
  architecture}, proposed in~\cite{RTW}. The goal of this architecture is to
{\em reduce the cost of deployment while achieving some performance
  constraints}. These constraints concern the event-to-sink
performance of the network~\cite{Akyildiz}, and thus require a {\em
  sufficient density of deployment} to correctly monitor the sensing
domain, and {\em efficient transport solutions} in order to provide
the detected information to the central unit.

The idea proposed in~\cite{RTW} assure some fraction of sensing capabilities
by the {\em transmit-only sensors}, who do not have receiver circuit,
hence are less energy consuming, and cheaper [Rabaey, ODonnell,
Sheng]. The remaining part of the sensing and the totality of the
information transport task is assured by regular sensors, also called
here {\em cluster-heads}.  In particular, they are responsible for
receiving information from transmit-only sensors, who send it blindly,
and forwarding it to sinks.

An immediate consequence of the assumption on the transmit-only
sensors is that the transmission traffic generated by them is
completely random, and thus, we have to admit that some part of the
detected information will be lost because of the collisions generated
at the receiver of the cluster heads. Note that arbitrarily increasing
the density of the transmit-only sensors and/or their traffic
intensity we may saturate the transport layer and thus make the
situation even worse.

In~\cite{RTW} the authors propose a simple heuristic of a packet admission
policy at a cluster-head in order to maximize the total number of
captured packet. In this paper, we define a detail mathematical model
of the hybrid network. Using this model we prove that the heuristic
from~\cite{RTW} is indeed optimal. We also derive the optimal policy that
maximizes the coverage region of a hybrid network, which was
previously unknown. Finally, using our model, we are able to quantify
the substantial savings obtained with the hybrid architecture.
 
In addition, the MAC model that we develop and call Erlang's loss
model with interference, might be of independent interest as adequate
for any non-slotted; i.e., unsynchronized, wireless communication
channel.

\paragraph*{Related work}
The only work on transmit-only sensor networks we are aware of
is~\cite{RTW}. An analysis of the event-to-sink performance of
standard network is
given in e.g.~\cite{Akyildiz}. Several works show a significantly lower
complexity and power consumption of transmiter over receiver circuit~\cite{Rabaey,ODonnell,Sheng}.

Note that we are {\em not} interested in this paper how to obtain the
sensor deployment that satisfies a correct monitoring of the
domain. Several results on this subject were already published;
e.g. using the explicit formula for the volume fraction of the
stochastic geometry Boolean model (see
eg.~\cite{LiuTowsley2003,DouseMannersaloThiran2004}) or asymptotic
formula, which can be applied when the density of nodes is large while
the sensing ranges are small; see~\cite{Janson86,Koskinen2004}.

The remaining part of this paper is organized as follows. In
Section~\ref{s.Assumptions} we describe the system assumptions. Next,
in Section~\ref{s.Analysis} we evaluate and optimize its performance.
Numerical examples are presented in Section~\ref{s.results}.
In Appendix we develop some details of the mathematical model
used to analyze the system.


\bsection{System assumptions}
\label{s.Assumptions}

Let us consider a network of {\em sensors} and {\em cluster-heads}. 
Sensors are simple sensing devices that are equipped  only with a
single transmitter. They are supposed to sense and periodically  send
information to the cluster-heads. 
Cluster-heads are more powerful (and more expensive) sensors. 
They are equipped with  a
receiver and a transmitter, and their special role is to collect
information from transmit-only sensors and forward it to a central server.
The network consists of a large number of sensors and
a much smaller number of cluster-heads. 
We want to analyze and optimize the {\em performance of the
information transport from sensors to cluster-heads}.

\bsubsection {Events and traffic}
\label{ss.EnT}
We assume that some events trigger transmissions at the  sensors 
randomly and independently of each other, with intensity $\lame$.
Two  scenarios are possible. 
In the first one, event is a time instants at which
a sensor decides to transmit information about  the actual state of the
sensed environment. There, $\lame$ is a system variable controlled by 
the designer. 
In the second scenario,  event is a time instant
at which a sensor senses some random excitation in its proximity
and transmits a report on it. A
local character of random excitations justifies the assumption of
independence between transmissions of different sensors.  
In this case, $\lame$ is an external parameter depending on type of events
we measure. 

Our analysis applies also to a situation when the excitation of the
medium is not local and persists for some time (like in an intrusion
detection problem). Then, the time and space scale of our spatial
throughput analysis corresponds to the duration of the excitation and
the region from which the sensors report about this excitation.  A
random, Aloha-type, back-off mechanism has to be implemented at
sensors, in order to avoid systematic packet collisions from other
sensors sensing the same excitation. The vent is a time instant
at which the back-off mechanism of a sensor makes it transmit; thus,
we find the first scenario in this interpretation. When the
excitation is over, the sensors may go to a sleep mode, or sense at
some smaller rate.

{\em A canonical example}, considered in this
paper,  is this of a channel with the throughput
of  1~MB/s at the physical layer, which is shared  by  sensors
homogeneously distributed with the density of  $\lambda_s=10$
sensors/m${}^2$.   
We consider the periods when the channel is actively used
by all the sensors which emit with the temporal intensity
$\lambda_e=1$ kbps. Such a relatively high intensity 
may be  reasonable to perform correctly during some
periods of  persistent excitations.

\bsubsection {Reception}
\label{ss.Rec} \label{ss.channel}
Since sensors are transmit-only, they cannot sense collisions, and
send packets blindly. 
The goal of cluster-heads is to receive these packets. We suppose 
that a packet is correctly  received if the 
SINR, empirically averaged over the reception duration,
is higher than some threshold. Otherwise, the packet  is lost. 
We consider a slow fading Gaussian channel channel model with {\em
  repetition coding with interleaving}. Repetition coding corresponds
to CDMA or UWB spreading. Interleaving means, that each bit is sent
through many symbols uniformly distributed over the duration of the
packet size.
Suppose that the receive is equipped with a matched filter (coherent
maximal ratio combiner).  Then, the standard analysis (see
e.g.~\cite[Section 3.2.1]{TseVis2005}) says, that there is a threshold
$\gamma$ on the SINR that should be respected in order to maintain the
link quality;
\begin{equation}\label{e.SIR}
\frac{|h|^2 P_{rec}}{W+1/M \sum_{j=1}^M I_j}\ge\gamma\,,
\end{equation}
where $W$ and $I_j$ is, respectively, the noise power and the power
received from interferers during the $j\,$ symbol, $P_{rec}$ is the
received power averaged over fading effects that is supposed to depend
only on the emitted power and the distance between the emitter and the
receiver.  Remark, that interleaving allows for the  empirical
  averaging of the interference $I$ in~(\ref{e.SIR}) over the packet
  duration.
We will interpret~(\ref{e.SIR}) as the SINR condition identifying the
successful reception of the packet at the MAC layer, given the link
fading value $h$ and the received interference power process
$\{I_j\}$. 

Commonly used fading model is Rayleigh fading where link fading $h$ is
represented with a circular complex Gaussian random variable. In this
case $|h|^2$ can be seen as a realization of some exponential random
variable (see e.g.~\cite[p. 50 and 501]{TseVis2005}).

\paragraph*{Cluster-head communication}
We also assume that cluster-heads have a reliable communication of a
higher rate to a central server, which does not interfere with the
sensor channel (e.g. can be wired, but not necessary).

\bsubsection {Synchronization and decoding}
\label{ss.Dec}
\begin{figure}[t]
\centerline{
     \includegraphics[width=0.8\linewidth]{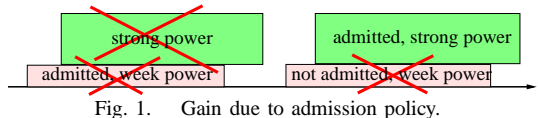}}
\vspace{-2.5ex}
\caption{\label{f.collision}
Gain due to admission policy.}
\vspace{-3ex}
\end{figure}

A cluster-head needs to synchronize to a packet sent by a sensor
before receiving it. In order to synchronize to a packet, the
cluster-head has to receive it with some minimum reception
power~\footnote{e.g. needed to have symbol-error rate during the
  preamble of $10\%$}.
If the packet reception power is higher than this, cluster-head starts
receiving and it continues receiving until the end of the
transmission. {\em If the transmission is lost because of the
  collision with another packet emission (interference), the error
  will be detected only at the end of the reception}. Moreover, {\em
  this interfering packet will be lost as well} since the cluster head
was not idle at its arrival epoch; cf. Figure~\ref{f.collision}.

In order to improve the efficiency, we introduce a {\em packet
  admission policy}. Once the cluster-head is synchronized to a
packet, it {\em may decide to receive or ignore the detected packet
  according to some packet admission policy} based, for example, on
the value of the received power.
This policy allows to ignore some weak packets so that the
cluster-head is more often available for stronger packets. The choice
of a particular policy depends on system design goals, described in
Section~\ref{ss.design_goals}.

\bsubsection{Sensor placement} Sensors in a given network realization
are fixed.  We show how to evaluate the performance of a given fixed
configuration of sensors served by one cluster head applying some
admission policy.
For performance optimization, we adopt a stochastic
approach. Namely, we optimize the system parameters with respect to the
average performance, taken over all possible spatio-temporal
configurations of packet emissions, which are driven by some
spatio-temporal Poisson point process.  We call this model a {\em
  Poisson-rain process of events}.  We argue in
Appendix~\ref{sss.Poisson-Rain} that this model --- which corresponds
to the situation where each event is associated with one sensor that
activates only once, when the event occurs, and leaves the system
afterward --- is a reasonable approximation of the packet traffic
generated be an arbitrary (also deterministic) repartition of sensors
on the plane, provided the density is large.  However, it is
important to keep in mind that this model is just an approximation,
that in reality the sensors are persistent, and that we can collect
and keep some information about each one.


\bsubsection {Design goals and performance metrics}
\label{ss.design_goals}

We consider several design goals, related to the transport of the
information from sensors to cluster-heads.
Our principal performance metric is the {\em spatio-temporal density
  $\rho(x)$ of the received information}. We define it as the mean
number of packets received from sensors, per second, from the surface
area $\md x$.

In order to obtain some desired density $\rho(\cdot)$
the system designer can influence the placement of nodes (at least
the density), transmission power, and frequency of transmissions. 
However, while the density of nodes can depend on the location, 
the transmission power and frequency is the same for all the sensors
as we suppose we cannot reconfigure each sensor separately once
it is placed. The main parameter that can be tuned 
in order to shape the information transport from sensors to the
cluster-heads is the packet admission policy applied by cluster heads.

In general, the network designer may be  interested in {\em maximizing 
the coverage} of some sensing domain or in {\em increasing the total    
throughput}. These contradictory  goals are realized by the policies 
respectively called {\em max-min} and {\em globally optimal}, which
are defined in Sections~\ref{ss.coverage} and~\ref{ss.throughput};

Finally, the goal of this paper is to propose a hybrid network that
will {\em minimize the cost of the network deployment}. We thus want
to show how much money can a designer save by combining wireless
tranceiver sensors with the cheap transmit-only ones. To that respect
we study the {\em economic optimization problem} in
Section~\ref{ss.cost}, which finds the right proportion of the two
types of devices that minimizes the cost of a network, under the
constraint of some desired level transport-aware effective coverage.


\bsection{Analysis}
\label{s.Analysis}

In this section we will analyze and optimize the performance of the
sensors-to-cluster-heads information transport in 
transmit only sensor network, whose system assumptions are described 
in Section~\ref{s.Assumptions}. 
We assume that the density of sensors
$\lambda_s$ is large enough and the sensing regions are small 
so as to  guarantee a good sensing-coverage of the domain with a sufficient
resolution. 

\bsubsection{Density of information}
First, we study the density of information 
received by a single cluster
head $\rho(x)$.
Remind that we define it as as the mean number of packets received by
the cluster-head per unit of time from the area $\md x$.  We will
model the traffic described in Section~\ref{ss.EnT} by a
spatio-temporal Poisson point process of events\footnote{We explain in
  Section~\ref{sss.Poisson-Rain} that this model, called there the
  {\em Poisson rain of events}, is a reasonable approximation of the
  process of packets transmitted be an arbitrary pattern of sensors
  (not necessarily Poisson) densely distributed on the plane.} (packet
transmissions) with intensity $\Lambda_s(\md x)\times\lambda_e\md t$,
where $\Lambda_s(\md x)=\lambda_s(x)\md x$ and $\lambda_e$ are,
respectively, the mean number of sensors placed at $\md x$, and the
temporal intensity of packet traffic sent by each sensor.
This Poisson rain of events (packets) is supposed to be received by
one cluster-head, whose behavior is described in
Sections~\ref{ss.Rec}--\ref{ss.Dec}.  Specifically, the cluster head
applies some admission policy $d(x)$, which is the probability that it
tries, given it is idle, to receive a given packet emitted from $\md
x$~%
\footnote{The cluster-head does not need to know the location of the
  receiver; it can apply some admission policy depending on the
  received power.}.
We assume that the admission decisions are taken independently of each
other and of anything else, and
thus the spatio-temporal process of {\em admissible} packets
is the Poisson process with intensity  $d(x)\lambda_s(x)\md
x\times\lambda_e\md t$. 

Inspired  be the  channel description of Section~\ref{ss.channel},
we assume that a given  admissible packet, arriving when the
cluster-head is idle,  is correctly received if some   
SINR, empirically averaged over the reception period $B$,
is higher than some threshold $\gamma$; cf.~(\ref{e.SIR}). 
The interference is created by all other
emissions  taking place at this time period and by some external
noise $W$. 
A detailed mathematical analysis of  the performance of the cluster-head
modeled by some  {\em Erlang's loss system with interference} and SINR
condition~(\ref{e.reception_OK}) is done
in Section~\ref{ss.SpatialErlang} under the assumption of {\em Rayleigh
fading}.  In what follows we summarize 
the results of this analysis.
First, we remind a general fact  that follows  from the Campbell
formula.
\begin{Prop}\label{p.rho}
The density of received information is equal to 
$\rho(x)=\lambda_e\lambda_s(x) d(x)p_{free}\,p_{rec}(x)$,
where 
$p_{free}$ is the probability that a typical admissible packet finds the
cluster head idle when it arrives
and $p_{rec}(x)$ is the conditional probability that the typical
admissible packet arriving from $\md x$ can be correctly received, 
given the cluster head stars receiving it.
\end{Prop}

Suppose that the cluster-head is located at the origin.
Denote by $\bar P$ is the emission power used be all sensors, 
by  $L(x)$ the power attenuation
function (path-loss) of the distance from $x$ to 0, and by $\calL_W$
the Laplace transform 
of the power $W$ of the external (white) noise; 
$\calL_W(\xi)=e^{-\xi  W}$ if this power is constant. 

\begin{figure}[t]
\centerline{
     \includegraphics[width=0.8\linewidth]{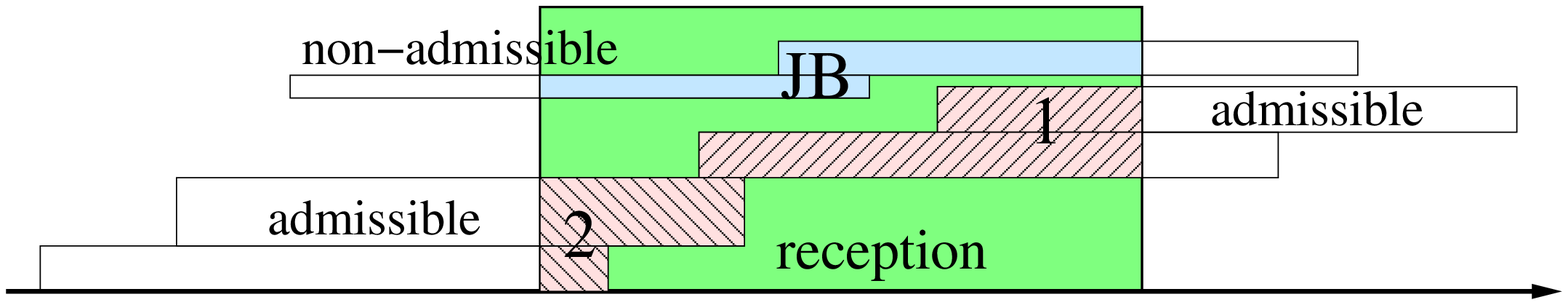}}
\vspace{-2ex}
\caption{\label{f.interference}
Three terms of interference $\calL_1,\calL_2,\calL_{J_B}$.}
\vspace{-3ex}
\end{figure}

For a given admissible packet received by the cluster-head, let 
$\calL_1,\calL_2,\calL_{J_B}$ be  the {\em Laplace transforms
of the interference averaged over the reception period}, generated
respectively, by: 
{\em admissible packets arriving when it is being received},
{\em admissible packets  that are being sent at its  arrival
   epoch}, {\em all non-admissible packets};
cf. Figure~\ref{f.interference}.
These Laplace transforms are explicitly given  
by formulas~(\ref{e.L1_rain}), (\ref{e.L2_rain}),
(\ref{e.JLT_rain})
with 
$\lambda=\lambda_e \int d(r)\,\lambda_s(x)\mcd x$
begin the total intensity of the admissible packets 
(the integral is taken over the whole domain of the network
deployment).
Denote $\gamma_x=\gamma /(\bar P L(x))$.
By Corollary~\ref{c.Erlang_rain}, we have the following result.
\begin{Prop} The Erlang acceptance probability is equal to 
$p_{free}=1/(1+\lambda B)$ and 
the conditional reception probability is equal to 
$p_{rec}(x)=\calL_W(\gamma_x)\calL_1(\gamma_x)
\calL_2(\gamma_x)\calL_{J_B}(\gamma_x)$. 
\end{Prop}

\begin{figure}[t]
\centerline{
\begin{minipage}[b]{0.58\linewidth}
     \includegraphics[width=1\linewidth]{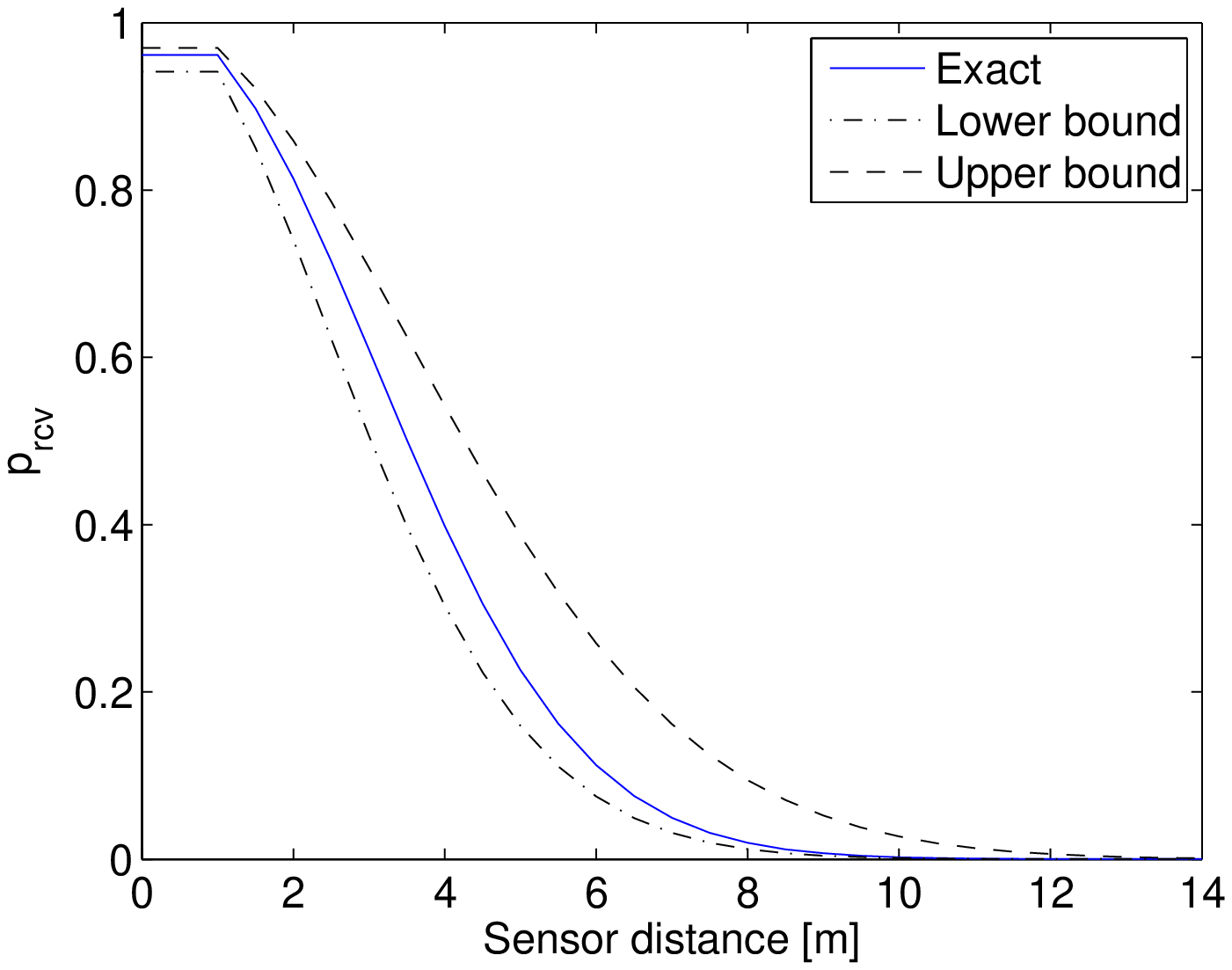}
\end{minipage} 
\hspace{-0.08\linewidth}
\begin{minipage}[b]{0.58\linewidth}
    \includegraphics[width=1\linewidth]{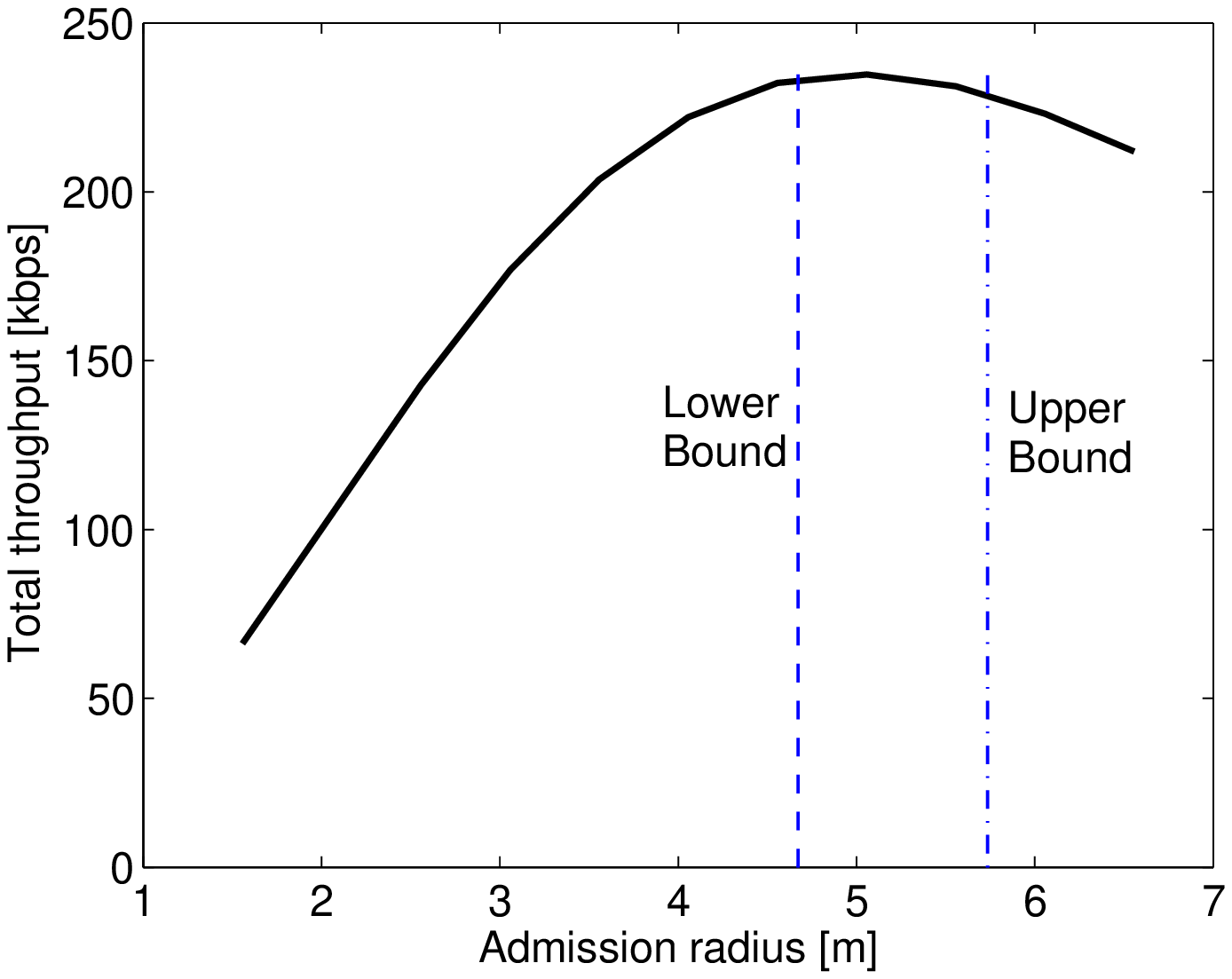}
\end{minipage}}
\caption{\label{f.bounds}
Exact value and approximations of $p_{rec}$ (left). Maximization of the
total intensity of information (right).}
\centerline{
\begin{minipage}[t]{0.58\linewidth}
     \includegraphics[width=1\linewidth]{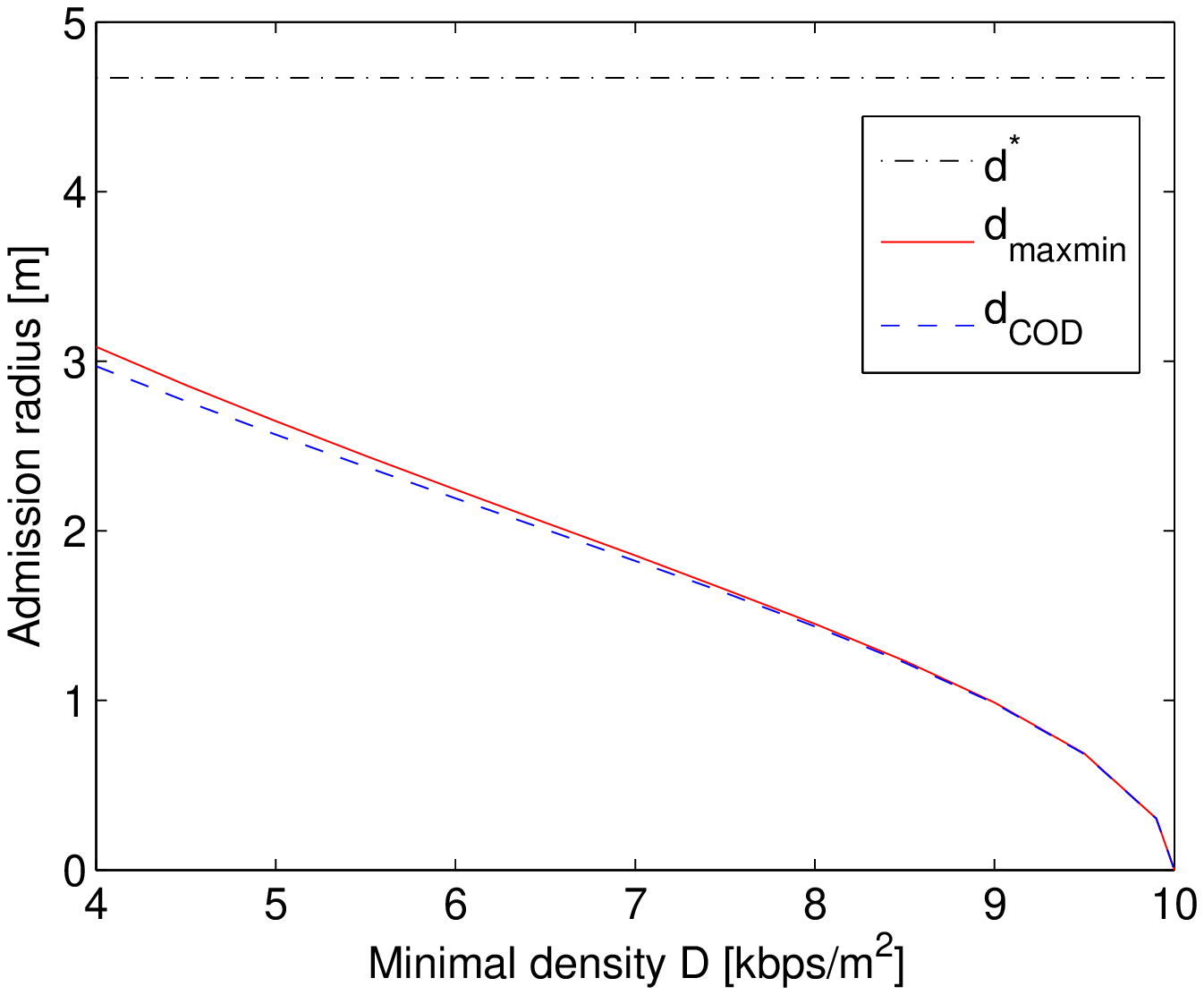}
\end{minipage} 
\hspace{-0.08\linewidth}
\begin{minipage}[b]{0.58\linewidth}
    \includegraphics[width=1\linewidth]{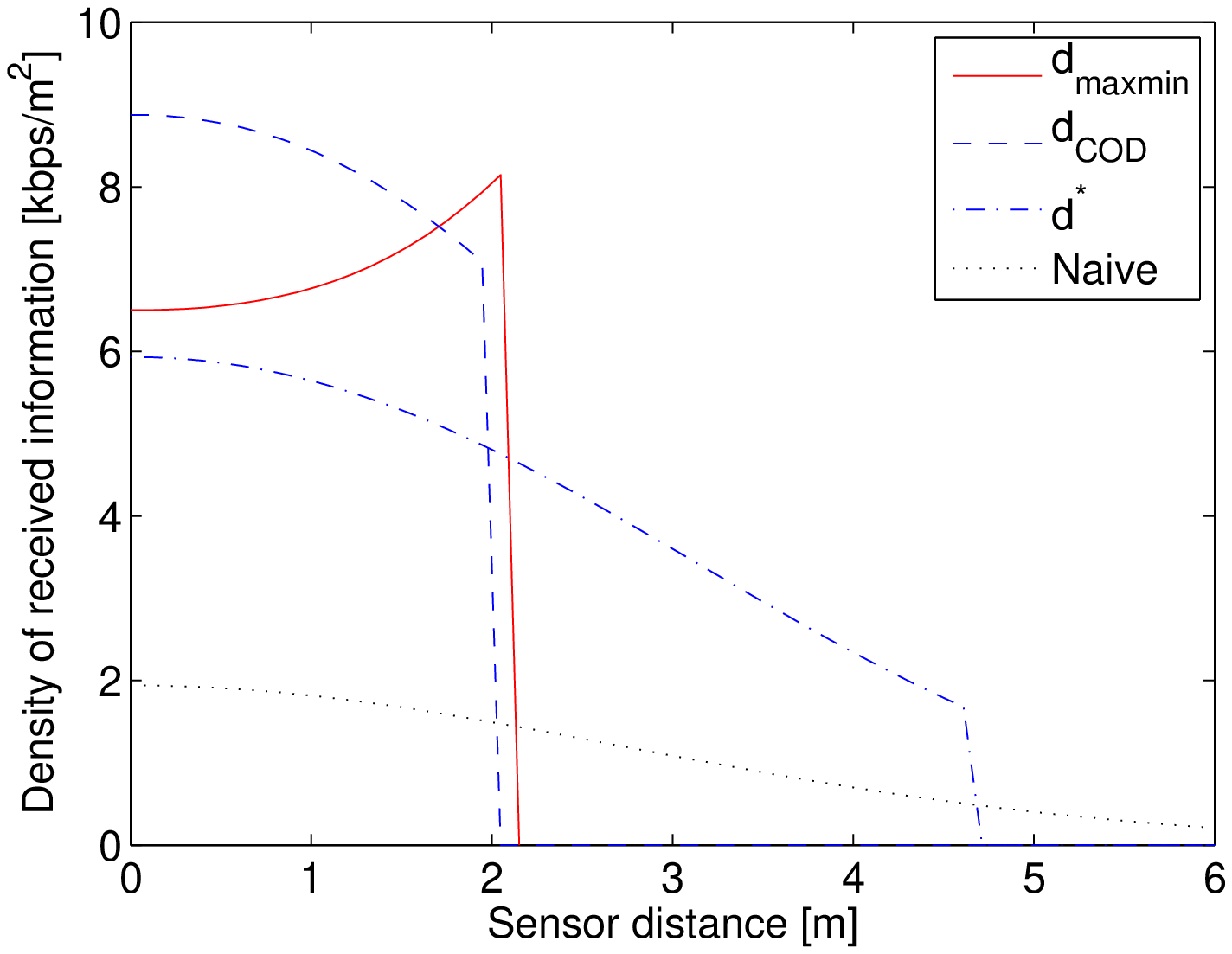}
\end{minipage}}
\caption{\label{f.policies}
Radii of different admission policies (left). The exact profiles of $\rho(x)$
for different admission policies (right) when applying different admission 
policies explicitly found using the lower-bound $\urho$ (this explains why 
the max-min policy does not give exactly $\rho(x)=const$ for $x$ within the 
admission distance).}
\centerline{
\begin{minipage}[b]{0.58\linewidth}
     \includegraphics[width=1\linewidth]{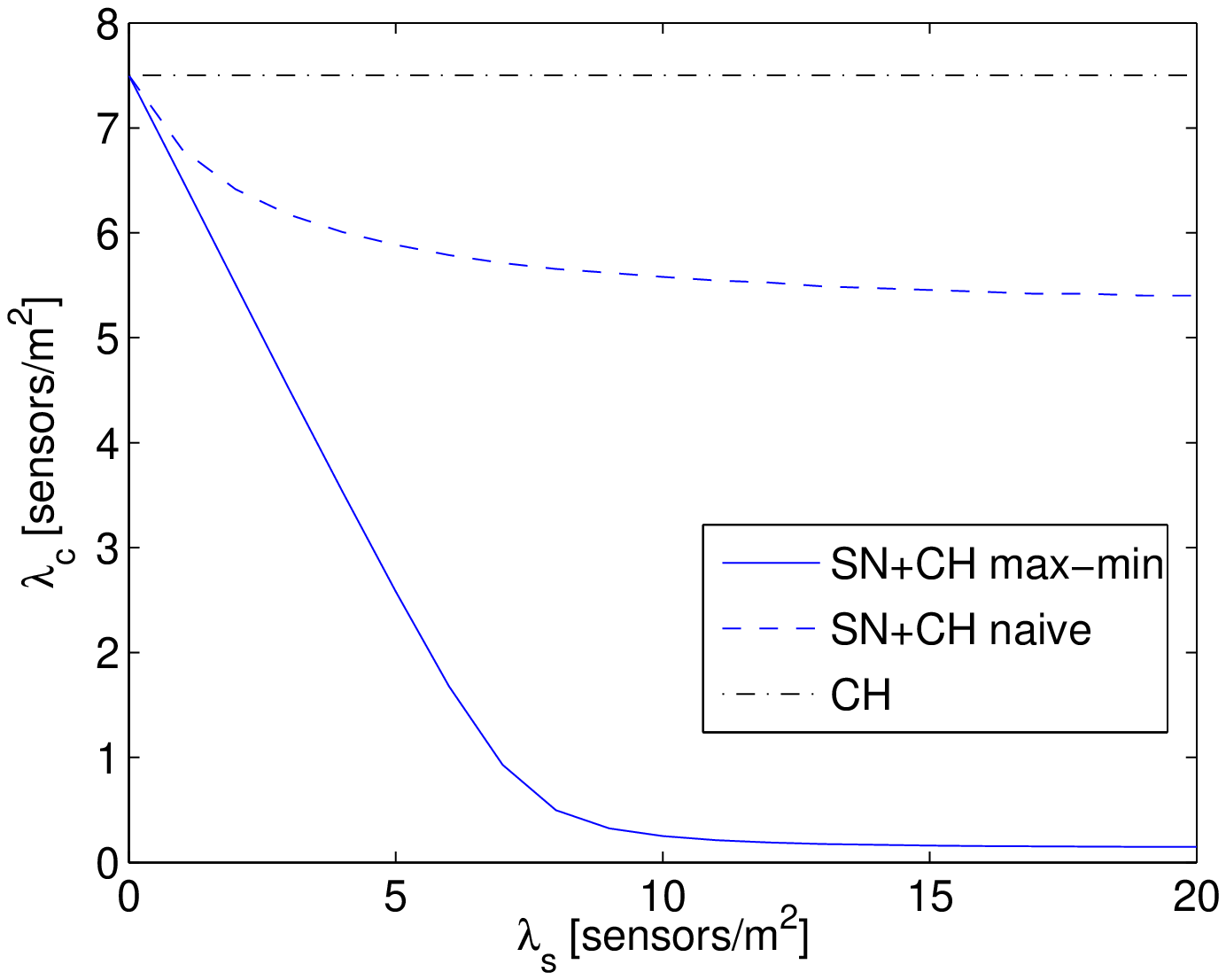}
\end{minipage} 
\hspace{-0.08\linewidth}
\begin{minipage}[b]{0.58\linewidth}
    \includegraphics[width=1\linewidth]{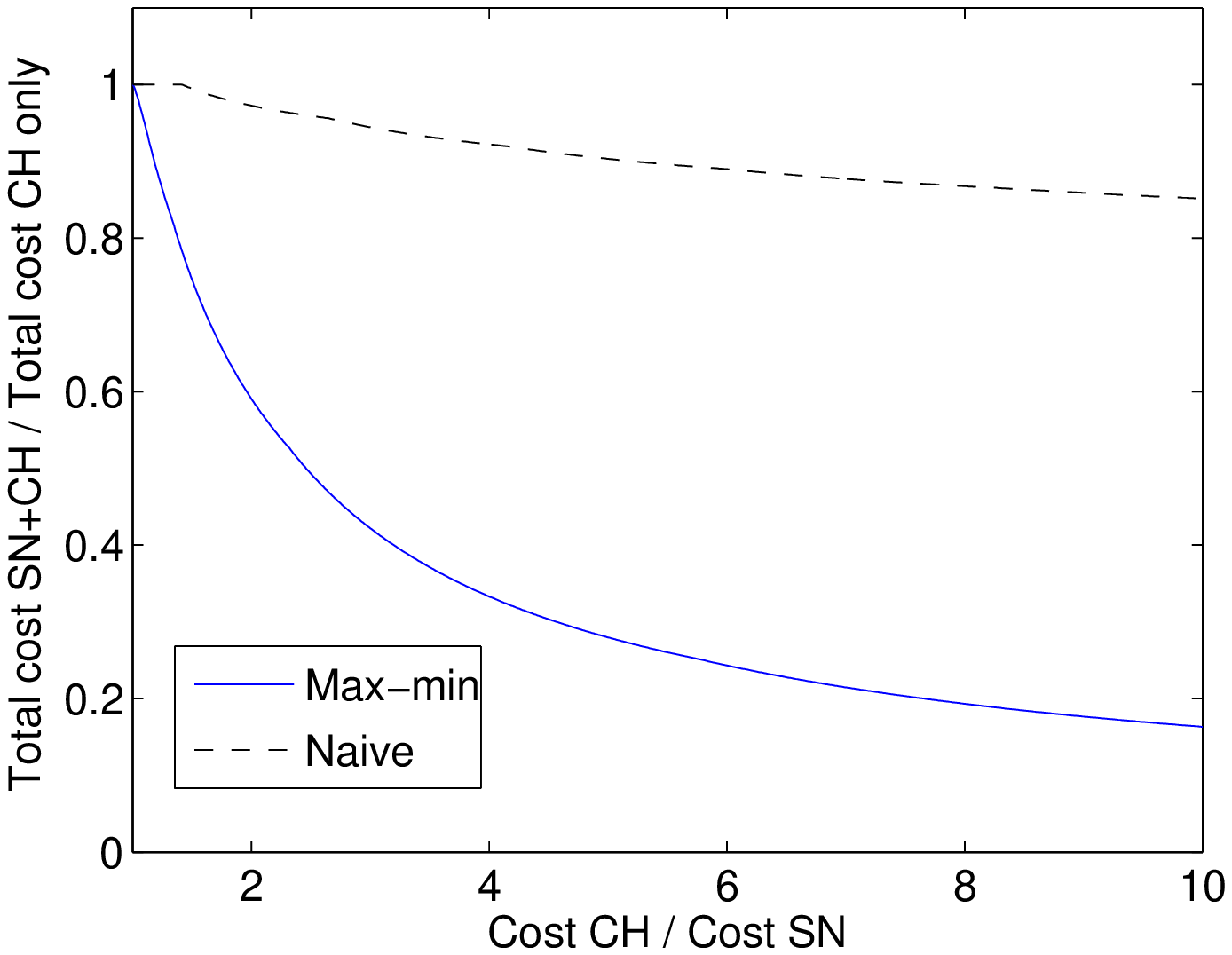}
\end{minipage}}
\caption{\label{f.cost}
Constitution of the hybrid network (cluster-heads CH and transmit-only
sensors SN) with minimal cost,
assuring $\min_x\rho(x)=0.75$  (right).
Network cost gain in function of the device cost per unit ratio (right).}
\vspace{-3ex}
\end{figure}

Lets denote 
\begin{equation}\label{e.calL}
\calL(\xi)=
\exp\biggl(\!\!\!-\lambda_eB\!\!\!
\int\!\!\Bigl(1-\frac{1}{\xi \bar P L(x)}
\log(1+\xi \bar P L(x))\,\Lambda_s(\md x)\Bigr)\!\!\biggr)\,.
\end{equation}
Corollary~\ref{c.bounds} gives two more explicit bounds on $p_{rec}(x)$.
\begin{Prop}\label{p.upop-rec}
We have $\up_{rec}(x)\!\le\!  p_{rec}(x)\!\le\! \op_{rec}(x)$,
where 
$\up_{rec}(x)=\calL_W(\gamma_x)\calL^2(\gamma_x)$,
$\op_{rec}(x)=\calL_W(\gamma_x)\calL(\gamma_x)$
and  $\calL$ is given by~(\ref{e.calL}).
\end{Prop}
Denote by $\urho(x), \orho(x)$, respectively, the lower and the upper
bound of  $\rho(x)$
obtained when $p_{rec}(x)$ in Proposition~\ref{p.rho}
is replaced by, respectively,  $\op_{rec}(x)$ and $\up_{rec}(x)$.
Note that the both $\op_{rec}(x)$ and $\up_{rec}(x)$ {\em do
not} depend on $d(\cdot)$ which makes the analysis of $\urho(x), \orho(x)$
easier (for the quality of  the  bounds see Figure~\ref{f.bounds}, (left)).

Before describing some optimal policies, we define a {\em naive
  policy} $d_{naive}(x)=\ind(x\in\bbD_0)$, where the set $\bbD_0$ is
fixed such that mean received power is large enough to receive
correctly the packet, given only external noise $W$ (no
interference)~\footnote{This may correspond to the
  successful synchronization to the packet}; i.e., $\bbD_0=\{x: \bar
PL(x)/W\ge \gamma\}$.

\bsubsection{Optimizing the  transport-aware  coverage}
\label{ss.coverage}
Knowing that the attenuation function $L(x)$ (and thus $p_{rec}(x)$)
typically decreases with the distance $|x|$ to the cluster-head, one
has to compensate it with an increasing density of sensors
$\lambda_s(\cdot)$ and/or a spatial admission policy $d(x)$.%

In this paper we suppose now that the sensors are already deployed
~\footnote{leving the the optimal deployment problem for future
work}  with
some given with density $\lambda_s(x)>0$ on some sensing domain
$\bbD$. 
We look for an admission policy $d(x)$, such that {\em any increase of
  the ratio $\rho(x)/D(x)$ on some set $\md x$ of positive Lebesgue's
  measure would be at the expense of decreasing of some already
  smaller ratio $\rho(y)/D(y)$} on some non-null set $\md
y\in\bbD$. The policy $\dmaxmin(x,D)$ realizing the above principle is
called {\em weighted max-min fair policy}, with weights $1/D(x)$.  It is
known that if $\dmaxmin(\cdot,D)$ exists then it is unique. For brevity
we will denote by $\dmaxmin(x)$ the max-min policy with 
equal weights ($D(x)=D$) (the policy  does not depend on the value of $D$).

We cannot exactly characterize the max-min fair policy for $\rho(x)$,
however, we can do this for some bounds.
Denote 
$\uI=\int_{\bbD} D(x)/\up_{rec}(x)\mcd x$.
Assume that the sensing domain $\bbD$ is compact, $D(x)$ continuous on $\bbD$
and denote 
$\uM=\max_{x\in\bbD}D(x)/(\lambda_s(x)\op_{rec}(x))$.
We define in the similar manner $\oI, \oM$ replacing
$\uprec(x)$ in the above formulas by $\oprec(x)$.


For a given policy $d(\cdot)$ denote by
$||d||_{\lambda_s}=\int_{\bbD}  d(x)\lambda_s(x)\mcd x$
the total spatial intensity of admissible packets under $d(\cdot)$.

\begin{Prop}\label{p.maxmin}
\begin{itemize}
\item 
The max-min fair policy $\udmaxmin(\cdot,D)$ for $\urho(x)$ on $\bbD$ 
exists if and only if $\uM<\infty$, it is equal to 
\begin{equation}\label{e.udmaxmin}
\udmaxmin(x,D)=\frac{D(x)}{\uM\lambda_s(x)\uprec(x)}\,
\end{equation}
and realizes $\urhomaxmin(x,D)=D(x)/(B\uI+\uM/\lambda_e)$.
Moreover, under $\udmaxmin(\cdot,D)$ we have $\rho(x)\ge\urhomaxmin(x,D)$. 
\item The max-min fair policy $\odmaxmin(\cdot,D)$ for $\orho(x)/D(x)$
on $\bbD$ 
exist if and only if $\oM<\infty$, it is given by~(\ref{e.udmaxmin})
with $\uM,\uprec$ replaced, respectively, by $\oM,\oprec$, 
and realizes $\orhomaxmin(x,D)=D(x)/(B\oI+\oM/\lambda_e)$.
Moreover,  there is no policy  $d(\cdot)$ for which
$\rho(x)\ge\orhomaxmin(x,D)$, with  
the strict inequality on some non-null set $\md x$. 
\end{itemize}
\end{Prop}
\begin{proof}
We consider the lower bound. The proof for the upper bound is analogous.
Suppose  that  $\uM<\infty$. Note that the function given by the
right-hand-side of~(\ref{e.udmaxmin}) in positive and not larger
than~1. Thus $\udmaxmin(\cdot,D)$  is a policy. 
Note also that for
$x_0=\maxarg_{x\in\bbD}D(x)/(\lambda_s(x)\up_{rec}(x))$ we have
$\udmaxmin(x_0,D)=1$. 
Assume now that for some
policy $d'(x)$ the respective ratio $\urho'(x)\ge D(x)
1/(B\uI+M/\lambda_e)$ and that the inequality is strict on some non-null
set $\md x$. It easy to show that then
$||d'||_{\lambda_s}>||\udmaxmin||_{\lambda_s}$ and thus
$\urho'(x_0)<\urho(x_0)$. This shows that $\udmaxmin$ is max-min fair.

Suppose now that $M=\infty$. Take any policy $d(\cdot)$. Note that
$\urho(x)/D(x)$ cannot be constant under this policy (there is no
such policy).
Thus, there exist $x_1,x_2$ such that $\urho(x_1)>\urho(x_2)$.
Note that we can slightly increase $d(x_1)$ and decrease $d(x_2)$ is such a
manner that $||d||_{\lambda_s}$ remains constant. This increases 
$\urho(x_2)$ without changing $\urho(x)$ for $x\not=x_1,x_2$. Thus,
$d(\cdot)$ is not a max-min fair. The remaining part of the  result
follows from Proposition~\ref{p.upop-rec}. 
\end{proof} 

\rem Suppose the cluster-head is to collect information
sent by sensors in a given compact domain $\bbD$ with some minimal
density $\rho(x)\ge D(x)$. The problem might be infeasible. However,
if it is, policy $\dmaxmin(x,D)$ satisfies the
constraint.

\begin{ex} 
\label{ex.1}
Consider a {\em uniform coverage}
$D(x)=D\times\ind(x\in\bbD)$ witght function. We might be interested in
maximizing the constant density $D$ given the domain $\bbD$. This is
achieved using $\dmaxmin$. 
Alternatively, we might be interested in maximizing the area of domain
$\bbD$ while providing some minimal density $D$. 
For example, for a homogeneous repartition of
sensor~$\lambda_s(x)=\lambda_s$ and distance-dependent path-loss
$L(x)=L(|x|)$ model, 
we  maximize the radius $R$ of the disk $\bbD=B(0,R)$ centered at 0, under
the contranit $\rho(x)\ge D$ for $x\in\bbD$.
Using Proposition~\ref{p.maxmin} one can find the 
solution $R=\uRmaxmin$ such that policy $\udmaxmin$ on
$\bbD = B(0,\uRmaxmin)$ satisfies $\rho(x) \geq \urho(x)
=D$ for all $x \in \bbD$. We illustrate this problem in
Section~\ref{s.results}.
\end{ex}

\bsubsection{Optimizing the total throughput}
\label{ss.throughput}
Consider now the problem of the maximization of the total weighted
intensity of received information $U=\int_{\bbD} \rho(x)/D(x)\mcd
x\,,$ where $D(x)>0$ are some arbitrary weights.

Denote by $\uU,\oU$, respectively, the lower and the upper bound of
the total weighted intensity of information
 obtained when $\rho(x)$ is replaced by
$\urho(x)$ and $\orho(x)$.
As previously, we can solve this global optimization problem for the
bounds  $\urho(x)$ and $\orho(x)$, and in this way approximate the
solution of the original problem.

Denote the following water-filling region
$\ubbD(\theta)=\{x\in\bbD: \lambda_s(x)\uprec(x)/D(x)>\theta\}$ and 
the constant 
$$\uthetas=\maxarg_\theta
\frac{\int_{\ubbD(\theta)}\lambda_s(x)\uprec(x)/D(x)\mcd x}%
{1+\lambda_eB\int_{\ubbD(\theta)}\lambda_s(x)\mcd x}\,.$$
We define in the similar manner $\obbD, \othetas$ replacing
$\uprec(x)$ in the above formulas by $\oprec(x)$.


\begin{Prop}\label{p.totalthr}
\begin{itemize}
\item The policy $\uds(x)=\ind(x\in\ubbD(\uthetas)$ maximizes  $\uU$.
Under this policy 
\begin{equation}\label{e.totalthr}
\uU=\uU^*=\frac{\lambda_e\int_{\ubbD(\uthetas)}
\lambda_s(x)\uprec(x)/D(x)\mcd x}%
{1+\lambda_eB\int_{\ubbD(\uthetas)}\lambda_s(x)\mcd x}\,.
\end{equation}
Moreover, under $\uds(\cdot)$ we have $U\ge \uU^*$. 
\item The policy $\ods(x)=\ind(x\in\obbD(\othetas)$ maximizes $\uU$.
Under this policy $\oU=\oU^*$ 
it is given by~(\ref{e.totalthr})
with $\ubbD,\uthetas$ replaced, respectively, by $\obbD,\othetas$, 
Moreover, there is no policy  $\d(\cdot)$ under which  
$U> \oU^*$. 
\end{itemize}
\end{Prop}
\begin{proof}
We consider the lower bound problem (proof for the upper bound is
analogous): 
maximize $\int_\bbD\lambda_sd(x)\uprec(x)/D(x)\mcd
x/(1-\lambda_eBA)$ under the constraints:
$A=\int_{\bbD}\lambda_s(x)d(x)\mcd x$, $0\le d(x)\le 0$.
We write the Lagrangian
\begin{eqnarray*}
\lefteqn{L(d,\theta,\mu_0,\mu_1)=\theta A+\int_{\bbD}\mu_1(x)\mcd x}\\
&&\hspace{-0.1\linewidth}+
\int_{\bbD} d(x)\Bigl(\frac{\lambda_s(x) \uprec(x)/D(x)}%
{1+\lambda_e BA}-\theta+\mu_0(x)-\mu_1(x)\Bigr)\mcd x\,.
\end{eqnarray*}
By the strong duality and the KKT conditions 
the optimal policy has the form of the indicator function 
$\uds(x)=\ind\bigl(\lambda_s(x)\uprec(x)/(D(x)(1+\lambda_e BA^*))
\le \theta^*\bigr)$ for some
$\theta^*,A^*$. The values of these constants are found by the
standard watter-filling  policy.  The remaining part of the  result
follows from Proposition~\ref{p.upop-rec}.
\end{proof}

\begin{ex} Consider equal weights $D(x) = D$, homogeneous
repartition of sensors  and  and distance-dependent path-loss model. Then
$\bbD(\uthetas) = B(0, \uRs)$ is a dics of radius $\uRs$~\footnote{In other
words, to maximize the total capacity it is optimal to receive only
packets whose received power is larger than some threshold.} We
illustrate this finding numerically in Section~\ref{s.results}.
\end{ex}


\bsubsection{Optimizing the network cost}
\label{ss.cost}
Suppose that one transmit-only sensor costs $C_s$, while a
transport-reliable 
sensor (with the same sensing functionality) costs $C_c$. 
Consider an architecture where the 
transport-reliable sensors act as cluster-heads considered previously in
this paper; call them cluster-heads.
Assume that information obtained (sensed) directly by cluster-heads
sensors is delivered to the central unit with probability one, while the
information obtained  by a transmit-only sensor located at $x$ 
is delivered there with
probability $p_{rec}(x-Z^*(x))$ where $Z^*$ is the location of the 
cluster-head nearest to $x$.

In order to formalize the problem of the economic optimization of the
proportion of the two types of devices, let us assume a regular
repartition of cluster heads on the plane.  A simple model consists in
taking them to be repartitioned on a regular, say, triangular grid
with 
some density $\lambda_c$. This means that $\lambda_c=4/(L^2\sqrt3)$,
where $L$ is the distance between two adjacent cluster-heads.  Note
that maximal distance to a nearest cluster head is equal to
$R_{\max}(\lambda_c)=4/\sqrt{\lambda_c3\sqrt{3}}$.  As as in the
previous section we model the traffic of packets sent be the sensors
to the cluster-heads (who act independently) by Poisson rain model of
events that is assumed to be stationary both in time and on the whole
plane $\ir^2$. To further simplify the model, we assume that at each
point $x$ in space at least one cluster-head has to achieve $\rho(x)
\geq D$. This is an upper bound on $\lambda_c$; in reality, a packete
that is lost by a cluster-head, may still be captured by another
cluster-head. However, this upper-bound is sufficient to numerically
demonstrate large savings of the hybrid approach.

Consider the following problem:
{\em minimize the cost of the network by unit area} 
$C=\lambda_sC_s+\lambda_cC_c$
{\em given some minimal intensity of received information}
$\lambda_e\lambda_c+\rho(R_{\max})\ge D$, where $\lambda_e\lambda_c$
is the density of information captured directly by the cluster-heads
and $\rho(R_{\max})$ is the lowest density of information that can by
obtained from the sensors given max-min (maximizing coverage)
admission policy.

In order to solve this problem, given $D$ and $\lambda_e,\lambda_s$,
we take the max-min policy $\udmaxmin$~(\ref{e.udmaxmin}) with
$D(x)=\ind(|x|\le R)$ and find the maximal radius $\uRmaxmin$, such
that the constant $\urhomaxmin$ obtained by the policy
$D(x)=\ind(|x|\le\uRmaxmin)$ on $B(0,\uRmaxmin)$ is equal to
$D-\lambda_e\lambda_c$ (cf. Example~\ref{ex.1}). 
Note by Proposition~\ref{p.maxmin} that for all $x$,  
$\rho(x)\ge\urho(x)\ge\urhomaxmin(\uRmaxmin)=D-\lambda_e\lambda_c$. 
This means that
taking $R_{\max}=\uRmaxmin$; i.e., $\lambda_c=4/(\uRmaxmin^23\sqrt3)$
is sufficient for 
$\lambda_e\lambda_c+\rho(R_{\max}(\lambda_c))\ge D$.  Having
calculated $\lambda_c=\lambda_c(\lambda_s)$ we express the cost of the
network $C=C(\lambda_c/\lambda_s)$ as the function the proportion
between the intensity of cluster-heads and the sensors.  Finally, we
look for its maximal value.


\bsection{Numerical results}
\label{s.results}

We will give now some numerical examples.  We consider the canonical
traffic scenario described in Section~\ref{ss.EnT} with SINR threshold
$\gamma=1$, path-loss model $L(x)=L(|x|)=\kappa |x|^{-\eta}$ with
$\kappa=10^{-5.5}$, $\eta=3.3$. 

\paragraph{Single cluster-head scenario}
In this part we consider a scenario with a single cluster-head at
0. Figure~\ref{f.bounds} (left) shows the the quality of
approximations given in Proposition~\ref{p.upop-rec}.

Next we compare maximum radii of different admission policies. In
addition to $\udmaxmin$ and $\uds$, defined in
Section~\ref{ss.coverage} and Section~\ref{ss.throughput}
respectively, we introduce the {\em coverage-optimal deterministic
  (COD) policy}. It is defined as $d_{COD}(x)=\ind(|x| \leq R_{COD})$
where $R_{COD}$ is the maximum radius such that $\rho(R_{COD}) \geq D$
under policy $d_{COD}$.

We can see from the results in Figure~\ref{f.policies} (left) that
$\oRmaxmin \approx R_{COD}$, hence that a deterministic policy with a
well-chosen radius provides almost as good coverage as the max-min
fair policy. In addition, we see from Figure~\ref{f.policies} (right)
that COD policy is more efficient. A more detailed discussion on a
tradeoff between efficiency and fairness is out of the  scope.

Figure~\ref{f.bounds} (right) shows the total intensity $U$ of
information obtained when the admission policy accepts all the packet
within a given radius. Optimal policy radius (maximizing the total
throughput) can be deduced from this plot. Two marked radii correspond
to policies $\ud^*, \od^*$.
We see that the policies $\ud^*, \od^*$ well approximate the optimal.
We also see on Figure~\ref{f.policies} (right) that maximizing capacity
requires admission region $\uRs$ that is much larger than $\uRmaxmin$
and  $R_{COD}$, having significantly smaller $\rho(x) <
\urho_{maxm}(x)$.

\paragraph{Economic optimization}
We now look at the economic aspects, described in
Section~\ref{ss.cost}. Figure~\ref{f.cost} (left) shows the required
density of cluster-heads in the hybrid network in function of the
density of transmit-only sensors, given the minimal value
$\min_x\rho(x)=\rho(R_{\max})=0.75$. Figure~\ref{f.cost} (right) shows the
network cost economization in function of the cluster-head/sensor unit
cost ratio. We can see that even when a price of a cluster-head is
only slightly higher than a price of a transmit-only sensor, we can
achive significant savings. On a contrary, using the naive policy,
cost savings are negligable.

\section{Implementation issues and concluding remarks}

In this paper, we analyzed hybrid sensor networks consiting of 
transceiving and transmit-only sensors. We presented a detailed
mathematical model of a physical and MAC layer of the network. 
Using this model we derived the optimal packet admission policies for
cluster-head that maximize coverage or total throughput. Also, using
the model we demonstrated how much the dollar-cost of a sensor network
can be decreased while maintaining the same network coverage. The MAC
model that we developed can also be used for any non-slotted wireless
communication channel.

In this work we did not discuss implementation details of the optimal
policies. However, we note that these policies can easily be
implemented, based only on the knowledge of packet received power (no
need to know channel attenuation function, sensor positions,
etc). One implementation is discussed in~\cite{RTW}, and can easily be
generalized to the policies proposed in this paper. We leave details for
future work.

Also, in this work we optimized only packet admission policy and not
the sensor density $\lambda_s(x)$, transmitting power
$\overline{P}(x)$, coding $\gamma(x)$ nor the transmission frequency
$\lambda_e(x)$, which may depend on particular sensors.
 However, we showed that even constrainted on
 the optimization of the admission policy only, 
we can achieve large savings and
maintaint architecture simple. Optimizing other parameters is left for
the future work.

\section*{Acknowledgment}
The authors are grateful to Fran{\c c}ois Baccelli for many helpful
discussions. 


\section*{Appendix: Mathematical modeling}\label{s.mathematical}
\setcounter{section}{1}
\setcounter{subsection}{0}
\setcounter{subsubsection}{0}
\renewcommand{\thesection}{\Alph{section}}
\setcounter{equation}{0}
\renewcommand{\theequation}{\thesection.\arabic{equation}}
\renewcommand{\theTh}{\thesection.\arabic{Th}}
\resetcounters

In this section we present mathematical models that are used to
analyze  the sensor network described in
Section~\ref{s.Assumptions}.
In particular,
%
%

\bsubsection{An Erlang's M/D/1/1 loss system with interference}
\label{ss.loss_system}
%
Assume a time homogeneous, independently marked  Poisson point process 
$\Phi=\{(T_n,(P_n,H_n))\}_{n=-\infty}^\infty$,
 where $T_n$ are
customer (packet)  arrival epochs and $(P_n,H_n)$ are 
independent, identically distributed (i.i.d.) marks, where 
$P_n\ge0$, $H_n\ge0$ can be interpreted as, respectively, the average
 (over fading   effects) power with which the 
$n\,$th packet arrives at the receiver and 
the actual fading state of its channel. (The randomness of 
 $\{P_n\}$ reflects different locations of transmitters  and
powers with which they emit packets, while the randomness of $\{H_n\}$
reflects the temporal variation of the channel conditions given fixed
 location of the transmitter and  emitted power.)
Lets denote by $\lambda$ ($0<\lambda<\infty$)
the intensity of $\Phi$; i.e., $T_n$ are
i.i.d. exponential r.v. with parameter~$\lambda$. 

We consider the following modification of the Erlang's loss policy.
Suppose that each arrival (i.e., packet) is admitted by the single
server of the system (i.e., starts being received by the receiver) if
this latter is idle at the packet's arrival epoch  
and rejected  otherwise. 
Admitted packets are received   during their duration time
$B$. However, packets  that are rejected by the receiver 
interfere during their emission time with the packets that  
are being received. Inspired by inequality~(\ref{e.SIR}),
with $H_nP_n=|h|^2P_{rec}$, we will say that 
the $n\,$th 
packet, given it is admitted by the receiver,
is correctly received  if the following inequality holds
\begin{equation}\label{e.reception_OK}
\frac{P_nH_n}{W+1/B\int_{T_n}^{T_n+B}I(t)\mcd t-P_nH_n}\ge\gamma \,,
\end{equation}
where $W$ is some nonnegative r.v. independent of $\Phi$,
$\gamma >0$ is   some constant,
$I(t)$ is the value of the  following temporal shot-nose process at time $t$
\begin{equation}\label{e.SN_temp}
I(t)=\sum_{i=-\infty}^\infty H_iP_i\ind_{[0,B)}(t-T_i)\,
\end{equation}
describing the total power received at time $t$ from all packets that
are being sent (including the power of the packet that is
being received; this is why we substract $P_nH_n$ from $I(t)$
in~(\ref{e.reception_OK}).  

Our goal is to calculate the 
frequency of the correct reception of packets; i.e.,
$\pi=\lim_{N\to\infty}\#\{\text{packets received among
  packets   
nr.\ } 1,\ldots,N \}/N$,
where $\#$ denotes the number.
Denote by $X(t)$ the indicator of the event that the receiver is busy at
time $t$ and  $X(-0)=\lim_{t\uparrow0}X(t)$.
Denote by $\delta_n$ the indicator of the event that the
inequality~(\ref{e.reception_OK}) holds.
Let $\Pr^0$ denote the  Palm probability given there is a customer
arrival at time $0$, and let $\EE^0$ denote the 
corresponding expectation. By Slivnyak's  theorem, under
$\Pr^0$ arrivals form the original stationary Poisson point process with an
extra arrival $(T_0,(P_0, H_0))$ added at time $T_0=0$ whose  mark
is independent and originally distributed.
Denote by $\Pr,\EE$, respectively, the stationary probability of $\Phi$
and its corresponding  expectation. The following results is a
consequence of the ergodic theorem (see
e.g.~\cite[Theorem~1.6.1]{BacBrem2003}).
\begin{Prop}\label{p.ErgodicTh}
The limit defining $\pi$ exists  $\Pr^0$ almost surely and 
$\pi=\EE^0[(1-X(-0))\delta_0]$.
\end{Prop}
In order to calculate $\pi$ we will first
characterize under $\Pr^0$ the distribution of the shot noise  process 
$I'(t)=I(t)-P_0H_0$ for $t\in[0,B)$
given the server is idle just before time 0 (i.e., $X(-0)=0$). 
It will be given in terms of the conditional joint Laplace
transform  
$\calL^0_{I'(t_1),\ldots,I'(t_n)|X(-0)=0}(\xi_1,\ldots,\xi_n)
=\EE^0\bigl[e^{-\sum_{i=1}^n I'(t_i)\xi_i}\big|X(-0)=0\bigr]$
evaluated for $\xi_i\ge0$
and any finite collection of time instants
$t_1,\ldots,t_n\in[0,B)$,
$n\ge1$.

\begin{Prop}\label{p.Cox}
Consider the Erlang's {\rm M/D/1/1} loss system with interference.
Then 
\begin{equation}\label{e.Erlang_form}
\Pr^0\{X(-0)=0\}=\frac{1}{1+\lambda B} 
\end{equation}
and for $t_1,\ldots,t_n\in[0,B)$, $n\ge1$, $\xi_1,\ldots,\xi_n\ge0$
\begin{eqnarray}\label{e.LT}
\lefteqn{
\calL^0_{I'(t_1),\ldots,I'(t_n)|X(-0)=0}(\xi_1,\ldots,\xi_n)}\\
&=&\int_0^\infty \lambda e^{-\lambda t}  \nonumber
\exp\Bigl[-\lambda \int_{-B}^B\Bigl(1-\ind_{(-t,0]}(s)\Bigr)\\
&&\times
\Bigl(1-\EE[e^{-\sum_{i=1}^n \xi_i H_iP_i\ind_{[0,B)}(t_i-s)}
\Bigr)\mcd s\Bigr]\mcd t\;. \nonumber
\end{eqnarray}
\end{Prop}

Note by the form of the above Laplace transform, 
that under $\Pr^0$ and given the server is idle just before
the arrival of the customer at $0$, the shot-noise process 
of interference $\{I'(t):0\le t<B\}$ is driven by a
non-homogeneous, double-stochastic  
Poisson process with intensity equal to $\lambda$ on the sum of the intervals
$(-B,-T]\cup(0,B)$ and 0 elsewhere, 
where $T$ is exponential random variable with
parameter $\lambda$. 

\begin{proof}
Note that (\ref{e.Erlang_form}) follows directly from the Erlang's
loss formula (see e.g.~\cite[equation~(81), p.~71]{Wolff1989}). 
In consequence, $\lambda/(1+\lambda B)$ is the intensity of the 
point process of arrivals of packets that are accepted by the receiver. 
In order to prove the remaining part of the proposition, 
lets define by $T(t)$ the time that elapsed form the last moment 
before $t$ when the receiver was busy; i.e., 
$T(t)=t-\sup\{s: s<t, X(s)=1\}$.
We will first show that under $\Pr^0$ and  given
$X(-0)=0$, the variable $T=T(0)$ is exponential with parameter
$\lambda$. 
Indeed, for $u>0$, by the Neveu exchange formula (see
e.g.~\cite[Section~1.3.2]{BacBrem2003}),  
\begin{eqnarray}
\Pr^0\{\,T(0)\ge u,X(-0)=0\,\}&=&\Pr^0\{\,T(0)\ge u\,\}
\nonumber\\
&&\hspace{-0.5\linewidth}=
\frac{1}{1+\lambda B}\, \EE^0_a
\biggl[\int_{(0,T_1^a]} \ind(T(t)\ge u)\,\Phi(\md t)\biggr]\,,\label{e.Neveu}
\end{eqnarray}
where $\EE^0_a$ corresponds to the  Palm probability, given a packet
arriving at time 0 is accepted by the receiver, $T_1^a$ is the next
arrival time after 0 of a packet accepted by the receiver, and the
integral with respect to $\Phi(\md t)$ denotes the sum over all
arrival times of the process $\Phi$.
It is easy to see that under $\EE^0_a$ we have
$T(t)=0$ for $t\in(0,B]$ and $T(t)=t-B$ for $t\in(B,T_1^a]$.
In the interval $(B,T_1^a]$ point process $\Phi$ has just one point,
namely $T_1^a$, and thus the integral in~(\ref{e.Neveu})
reduces to $\ind(T(T_1^a)\ge u)=\ind(T_1^a-B\ge u)$.
The distribution of the points of $\Phi$ in $(0,\infty)$ is not
influenced by the condition that the server is idle just before $0$ and
that there was an arrival at $0$.
Thus, under $\EE_a^0$, as well as under $\EE$, it is equal to the
distribution of points of the  the original Poisson point process.  
Thus, due to the lack of memory of the exponential inter-arrival r.v., 
the variable $T_1^a-B$ is  exponential with parameter $\lambda$, which
completes this part of the proof. 

Note now, that the packets which contribute to $I'(t)=I(t)-P_0H_0$ for
$t\in[0,B)$, 
given $T(0)=T$ and $T_0=0$, arrive only during the time intervals 
 $(-B,-T)\cup(0,B)$. (Note that $\Pr^0($ there is an  arrival at $-T)=0$
 if $T>0$.) Note also, that this set is disjoint with the set
$\calS=[T^*,-B-T(0)]\cup[-T(0),0]$,
where $T^*=\sup\{T_n:T_n<T(0)-B, T_{n+1}<T(0)-B, T_{n+1}-T_n>B\}$.
This latter set is a random stopping set (i.e., for a given
realization $\phi$ of the point process  $\Phi$, the set $\calS(\phi)$
is invariant with respect to any modification of the points of $\phi$
in $\ir\setminus\calS(\phi)$; see e.g.~\cite{Zuyev2006}).
Also, $X(-0)$ and $T(0)$ depend only on the configuration of points of
$\phi$ in $\calS(\phi)$. Thus, 
by the strong Markov property of the Poisson point process
(see~\cite{Zuyev2006}), given  $X(-0)=0, T=T(0)$, the distribution of 
arrivals  in $(-B,-T)\cup(0,B)$ is equal to 
the original distribution of points of the independently marked Poisson point
process  $\Phi$ taken on this region, and hence~(\ref{e.LT}) holds.
This completes the proof.  
\end{proof}

Before we  give an explicit formula for the frequency $\pi$ of the correct
reception of packets for our loss system with interference in the case
of Rayleigh fading, we will calculate the conditional Laplace transform
of the integrated shot-noise $I'_B=1/B\int_0^B I'(t)\mcd t$ 
given $\Pr^0$ and $X(-0)=0$; i.e.,
$\calL^0_{I'_B|X(-0)=0}(\xi)=
\EE^0\bigl[e^{-\xi I'_B}\big|X(-0)=0\bigr]$ for $\xi\ge0$.

\begin{Prop}\label{p.IB}
Suppose $\{H_n\}$ are exponentially distributed with mean~1
and $\{P_n\}$ are independent of $\{H_n\}$.
Then $\calL^0_{I'_B|X(-0)=0}(\xi)=\calL_1(\xi)\calL_2(\xi)$,
where 
\begin{eqnarray}\label{e.I1}
\calL_1(\xi)&=&
\exp\biggl(-\lambda B\Bigl(1-\EE\Bigl[\frac{1}{\xi P}
\log(1+\xi P)\Bigr]\Bigr)\biggr)\,,\\
\calL_2(\xi)&=& \exp(-\lambda B)\label{e.I2}\\
&\times&\!\!\!\!\!\biggl(1+\lambda B\int_0^1 
\exp\biggl(\lambda B\EE\Bigl[\frac{1}{\xi P}
\log(1+\xi P t)\Bigr]\biggr)\mcd t\biggr)\,.
\nonumber
\end{eqnarray}
\end{Prop}

\begin{proof}
Note first, that the integrated shot noise is also a shot noise type
variable. Indeed, $I'_B=1/B\int_0^B\sum_{i\not=0}
H_iP_i\ind_{[0,B)}(t-T_i)=\sum_{i\not=0} H_iP_i V(T_i)$,
where $V(t)$ is equal to $B-|t|$ for $|t|\le B$ and 0 elsewhere. 
By the Proposition~\ref{p.Cox}, it suffices to calculate the Laplace
transform of $I'_B$  driven by independently marked double-stochastic  
Poisson process with intensity $\lambda$ on the sum of the intervals
$(-B,-T]\cup(0,B)$ where $T$ is exponential random variable with
parameter $\lambda$. 
Denote 
\begin{eqnarray}\label{e.I11}
I^1_B&=& \sum_{i\not=0} H_iP_i V(T_i)\ind_{(0,B)}(T_i)\,,\\
I^2_B&=& \sum_{i\not=0} H_iP_i V(T_i)\ind_{(-B,-T]}(T_i)\,.
\label{e.I22}
\end{eqnarray}
Applying the general formula of the Laplace transform of the
independently marked Poisson point process (see e.g.~\cite{SKM95}))
we get
\begin{equation}\label{e.I1B}
\EE^0[e^{-\xi I^1_B}|X(-0)=0]
=\exp\biggl(-\lambda\!\!\!\int_0^B\!\!\!\Bigl(1-\EE\Bigl[e^{-\xi H P
V(t)}\Bigr]\Bigr)\mcd t\biggr)
\,,
\end{equation}
where $H, P$ are independent, generic r.v's  for $\{H_n\}$ and
$\{P_n\}$. Integrating with respect to $t$ and evaluating expectation
with respect to the exponential r.v. $H$ we obtain  
$\EE[e^{-\xi I^1_B}]=\calL_1(\xi)$.
In order to calculate $\EE^0[e^{-\xi I^2_B}|X(-0)=0]$,
we condition on $T=T(0)$ and 
we use  similar arguments, with integral $\int_0^B$ in~(\ref{e.I1B})
replaced be $\int_{-B}^{-T}$
(note that this integral is null if $T>B$). 
Similarly as for $I^1_B$ and then by 
integration with respect to the law of exponential $T$ we obtain 
$\EE^0[e^{-\xi I^2_B}|X(-0)=0]=\calL_2(\xi)$.
Obviously $I'_B=I^1_B+I^2_B$, and  variables $I^1_B,I^2_B$ are
independent, 
because they are driven by disjoint regions of
the underlying Poisson point process. Thus
$\calL^0_{I'_B}(\xi)=\calL_1(\xi)\calL_2(\xi)$, which completes the
proof.
\end{proof}

Now we are able to give the main result  of this Section --- an
Erlang's type formula.

\begin{Prop}\label{p.Erlang}
Consider the Erlang's {\rm M/D/1/1} loss system with interference.
Suppose that $\{H_n\}$ are exponential r.v's independent of $\{P_n\}$
and  lets denote by $\calL_W(\xi)$ the Laplace transform of $W$.
Then 
$$
\pi=\frac{1}{1+\lambda B}
\EE\Bigl[\calL_W(\gamma /P_0)\calL_1(\gamma /P_0)
\calL_2(\gamma /P_0)\Bigr]\,,
$$
where the expectation is taken with respect to the random
variable $P_0$.
\end{Prop}

\begin{proof}
By Proposition~\ref{p.ErgodicTh} $\pi$ is equal to 
\begin{eqnarray*}
\lefteqn{\EE^0[(1-X(-0))\delta_0]}\\ 
&&\hspace{-0,1\linewidth}=\Pr^0\{\,X(-0)=0\,\}\,\EE^0[\delta_0|X(-0)=0]\\
&&\hspace{-0,1\linewidth}= \Pr^0\{\,X(-0)\,\}\,
\Pr^0\Bigl\{\,H_0\ge \gamma W(W+I'_B)/P_0\Big| X(-0)=0\,\Bigr\}\\
&&\hspace{-0,1\linewidth}=\Pr^0\{\,X(-0)=0\,\}\,
\EE^0\Bigl[\exp\Bigl(-\gamma (W+I'_B)/P_0\Bigr)\Big| X(-0)=0\Bigr]
\end{eqnarray*}
because $H_0$ is exponential with mean 1 and independent of everything
else. Conditioning on $P_0$, noting that $W$ is independent of $I'_B$
and using Propositions~\ref{p.Cox} and~\ref{p.IB} we obtain the
result.
\end{proof}

Lets introduce now to the  Erlang's loss model an  
additional (external) stationary, ergodic process $J(t)$ of interference, 
independent of $W$ and $\Phi$. 
(For example, one can think of $J(t)$ as of the  interference created by
emitters transmitting packets that are not supposed to be received by
our receiver due to some random, independent admission policy;
cf. Section~\ref{ss.SpatialErlang}). 
Lets say that the $n\,$th 
packet of $\Phi$, given it is admitted by the receiver,
is correctly received  if 
to the inequality~(\ref{e.reception_OK}) holds
with the term $1/B\int_{T_n}^{T_n+B}J(t)\mcd t$ added to the denominator.
Denote by $\delta'_n$ the  indicator of this event.
and $\pi'=\EE^0[(1-X(-0))\delta'_0]$.
Denote by $J_B=\int_0^BJ(t)\mcd t$
and its Laplace transform by $\calL_{J_B}(\xi)=\EE[e^{-\xi J_B}]$.
We have the following straightforward extension of the
Proposition~\ref{p.Erlang}.
\begin{Cor}\label{c.ErlangJ}
Consider the Erlang's {\rm M/D/1/1} loss system with the external
interference $J(t)$. Under the same assumptions as in
Proposition~\ref{p.Erlang} the frequency of the correct reception of
packets exists $\Pr^0$ almost surely and is equal to 
$$\pi'=
\frac%
{\EE\bigl[\calL_W(\gamma /P_0)\calL_1(\gamma /P_0)
\calL_2(\gamma /P_0)\calL_{J_B}(\gamma/P_0)\bigr]}%
{1+\lambda B}\,.
$$
\end{Cor}

\bsubsection{Sensors on the plane}
\label{ss.SpatialErlang}
In this section we assume that the packets are emitted from different
locations of  the plane $\ir^2$ and, assuming some form of the attenuation
function, we will obtain a particular form of the distribution of
the powers  $\{P_n\}$ received at the origin, where the receiver is
supposed to be  located.
(Note that this  distribution was not specified in the previous
section.) We will also assume some packet admission policy.

\paragraph*{Attenuation function}
Suppose that the signal transmitted 
with some power $\bar P$ at the location $x$ is attenuated 
on the path to the receiver located at 0 (on average over fading effects) 
by the factor $L(x)>0$.; i.e., 
the average received power  is equal to $\bar P L(x)$. 

\paragraph*{Spatial policy of packet admissions}
\begin{figure*}[!t]
\normalsize
\begin{eqnarray}\label{e.L1_rain}
\calL_1(\xi)&=&
\exp\biggl(-\lambda B+ \lambda_e B
\int
\frac{d(x)}{\xi \bar P L(x)}
\log(1+\xi \bar P L(x))\,\Lambda_s(\md x)\biggr)\,,\\
\calL_2(\xi)&=& \exp(-\lambda B)
\biggl(1+\lambda B\!\! \int_0^1\!\! 
\exp\biggl(\lambda_e B
\int \frac{d(x)}{\xi \bar P L(x)}
\log(1+\xi \bar P L(x) t)\,\Lambda_s(\md x) \biggr)\mcd t\biggr)\,,
\label{e.L2_rain}\\
\calL_{J_B}(\xi)&=&
\exp\biggl(
-2\lambda_e B\!\!
\int\!\!(1-d(x))\Bigl(1-\frac{1}{\xi \bar P L(x)}
\log(1+\xi\bar P L(x)) \Bigr)\Lambda_s(\md x)\biggr)\,.
\label{e.JLT_rain}
\end{eqnarray}
\hrulefill
\vspace*{0pt}
\end{figure*}

Suppose that packets are emitted from different locations
of the plane $\ir^2$.
Suppose moreover, that the receiver located at the origin 
adopts the following {\em spatial admission policy}. 
Depending on emission location  $x$,
it accepts the packet, independently of everything else, with
probability  $d(x)$ (and starts receiving it, provided it is idle),
where $0\le d(x)\le 1$ is a given  function of location~$x$. 

\bsubsubsection{Poisson rain model of events}
\label{sss.Poisson-Rain}
Consider a spatio-temporal Poisson process 
$\{(X_n,T_n)\}$ where $X_n\in\bbD\subseteq\ir^2$, $T_n\in\ir$,
with intensity measure $\Lambda_s(\md x)\times \lambda_e\mcd t$.
The coordinates  of the  point $(X_n,T_n)$ denote, respectively,
the location of a packet  emission and the time it starts.
(One can think of emitters being born 
at locations $X_n$ and time $T_n$ just to emit one packet at this
moment; after the transmission of this packet the emitter disappears.) 
We assume that the points $\{(X_n,T_n)\}$  are 
independently marked by  i.i.d. exponential (with mean 1)
random variables
$H_n$ modeling the fading conditions during the transmission $n$.
Moreover, assuming some admission policy $d(\cdot)$ we suppose that
the points are further marked by i.i.d Bernoulli variables $U_n$
describing the admission status of the packets; 
i.e., $\Pr\{\,U_n=1\,\}=1-\Pr\{\,U_n=0\,\}=d(X_n)$,
where $U_n=1$ marks an admissible packet.

We call the  marked Poisson point process 
$\Psi=\Psi_d=\bigl\{\bigl((X_n,T_n),(H_n,U_n)\bigr)\bigr\}_n$,
the  {\em Poisson rain of events with a given
spatial admission policy $d(\cdot)$}.
We consider  $\Psi$  as the input to the 
Erlang's loss system with interference described in
Sections~\ref{ss.loss_system}. Specifically, we define
$P_n=\bar P\,L(X_n)$ 
and take the admissible packet transmissions 
$\Phi=\{(T_n,(P_n,H_n)):U_n=1\}_n$ as the input to the system,
whereas the total received power from non-admissible packet transmissions
\begin{equation}\label{e.J}
J(t)=\sum_n U_nH_n\bar P L(X_n)\ind_{[0,B)}(t-T_n)\,,
\end{equation}
as the external interference.
Denote by 
\begin{equation}\label{e.lambda_rain}
\lambda=\lambda_e\int d(x)\,\Lambda_s(\md x)\,
\end{equation}
the  (temporal) arrival intensity of Poisson process of the packets
admissible according to the spatial policy $d(\cdot)$.
The following consequence of Corollary~\ref{c.ErlangJ}
gives the  Erlang's type formula for the Poisson rain model.
\begin{Cor}\label{c.Erlang_rain}
Consider the Erlang's loss system $M/D/1/1$ driven by the
Poisson rain of events $\Psi$ on some domain $\bbD$ with 
spatial admission policy $d(\cdot)$.
Assume that $\lambda$ given by~(\ref{e.lambda_rain}) is finite.
Then, the fraction $\pi'=\pi'(x_0)$ of {\em admissible} packets correctly
received from a location 
$x_0\in\ir^2$, given there is an emitter located there, is given
by Corollary~\ref{c.ErlangJ}, with constant $P_0=\bar P L(x_0)$,
and $\calL_1,\calL_2,\calL_{J_B}$ given, respectively,
by~(\ref{e.L1_rain}), (\ref{e.L2_rain}), (\ref{e.JLT_rain}), where
$\lambda$ is given by~(\ref{e.lambda_rain}) and 
the integrals are taken over $\bbD$.
\end{Cor}
\begin{proof}
Note that the distribution of the received power is equal to
$\Pr\{\, P\le a\,\}=
\int d(x)\ind\bigl(\bar P L(x)\le
a\bigr)\,\Lambda_s(\md x)/\int d(y)\,\Lambda_s(\md y)$,
which  is correctly defined since we assume  $\lambda<\infty$.
Then, formulas~(\ref{e.L1_rain}), (\ref{e.L2_rain}) follow,
respectively,  from~(\ref{e.I1}), (\ref{e.I2}). 
Next, note that $\Phi$ and $J(t)$ are independent; this is a consequence of
the independent thinning of the Poisson process of all packet emissions.
Moreover, the integrated interference $J_B$,
given by the formula
$J_B=\sum_n(1-U_n))H_n\bar P L(X_n) V(T_n)$
is a shot-noise 
type random variable (cf. the proof of Propositon~\ref{p.IB}). 
Its  Laplace transform $\calL_{J_B}(\xi)$ is known explicitly and
given  by~(\ref{e.JLT_rain}).
\end{proof}

\bsubsubsection{Fixed arbitrary locations of emitters}
Suppose emitters are fixed and located at  $\{X_i\}$.
This case can be seen as a special case of the Poisson-rain of
packets, with purely atomic spatial density measure
$\Lambda_s(D)=\#\{X_i:X_i\in D\}$. Then the 
integrals $\int(\dots)\mcd x$ 
in formulas~(\ref{e.L1_rain})-(\ref{e.JLT_rain})
take form of the respective sums $\sum_{X_i}(\dots)$. 
Moreover, if  the spatial repartition of $X_i$ is sufficiently dense, 
then these atomic measures can be reasonably ``smoothed''
leading to approximative integral formulas. 
In particular, if the repartition is dense and uniform (in empirical
sense) then the sums can be approximated  by integrals with respect to the
Lebesgue's measure $\Lambda(\md x)=\lambda_s\mcd x$ with
$\lambda_s=\#\{X_i\}/|\bbD|$   where $|\bbD|$ is the surface of $\bbD$.

\bsubsubsection{Bounds}
In this section we will give some simple bounds for the frequencies
of successful receptions of packets. 
Denote $\gamma_x=\gamma /(\bar P L(x))$.
\begin{Cor}\label{c.bounds}
Under the assumptions of Corollary~\ref{c.Erlang_rain}, we have
$$\frac{\calL_W(\gamma_{x_0})(\calL(\gamma_{x_0}))^2}{1+\lambda B}\le
\pi'(x_0)
\le \frac{\calL_W(\gamma_{x_0})\calL(\gamma_{x_0})}%
{1+\lambda B}\,,
$$
where $\calL$ and $\lambda$ are  given by~(\ref{e.calL})
and~(\ref{e.lambda_rain}), respectively.
\end{Cor}
\begin{proof}
Note first that $\calL_2(\xi)\ge\calL_1(\xi)$. This can be verified
directly comparing formulas~(\ref{e.I1}) and~(\ref{e.I2}), but a
simple probabilistic argument can be used as well; remind that 
that $\calL_1(\xi)$ is the Laplace
transform of $I^1_B$ given by~(\ref{e.I11}), whereas $\calL_2(\xi)$ is
the Laplace transform of $I^2_B$ given by~(\ref{e.I11}).
Than, the lower bound
of Corollary~\ref{c.bounds} follows immediately from~(\ref{e.L1_rain})
and~(\ref{e.JLT_rain}). 
In order to get the upper bound, it is enough to observe that
$\calL_2\le1$ and to take $\calL_{J_B}$ with factor $2$ in the exponent
of ~(\ref{e.JLT_rain}) replaced by~1.
\end{proof}
Note that the upper bound in Corollary~\ref{c.bounds} consists in
taking no interfering arrivals before reception of a given packet,
whereas the lower bound consists in assuming the unconditional Poisson
process of such arrivals ($T=0$ in the proof of Proposition~\ref{p.Cox}).


\bibliographystyle{IEEEtran}
\bibliography{biblio}

\end{document}